\documentclass[twocolumn]{aastex6}
\usepackage{epstopdf}
\usepackage{amsmath,array,graphicx}

\usepackage{array,graphicx,tabularx,threeparttable, capt-of}
\usepackage{booktabs}
\usepackage{pifont}
\usepackage{multirow}
\usepackage{natbib}

\newcommand\lamb{\lambda_{\rm B}}
\newcommand\lamd{\lambda_{\rm D}}
\newcommand\fb{f_{\rm b}}
\fullcollaborationName{The Friends of AASTeX Collaboration}

\begin{document}


\title{Magnetic reconnection may control the ion-scale spectral break of solar wind turbulence}
%
%


\author{Daniel Vech\altaffilmark{1}, Alfred Mallet\altaffilmark{2,3}, Kristopher G. Klein\altaffilmark{1,4} and Justin C. Kasper\altaffilmark{1},}

\affil{$^{1}$Climate and Space Sciences and Engineering, University of Michigan, Ann Arbor, MI 48109, USA, $^{2}$ Space Science Center, University of New Hampshire, Durham, NH 03824, USA, $^{3}$Space Sciences Laboratory, University of California, Berkeley CA 94720, USA, $^{4}$Lunar and Planetary Laboratory, University of Arizona, Tucson, AZ 85719, USA; dvech@umich.edu}
\begin{abstract}

The power spectral density of magnetic fluctuations in the solar wind exhibits several power-law-like frequency ranges with a well defined break between approximately 0.1 and 1 Hz in the spacecraft frame. The exact dependence of this break scale on solar wind parameters has been extensively studied but is not yet fully understood. Recent studies have suggested that reconnection may induce a break in the spectrum at a ``disruption scale"  $\lamd$, which may be larger than the fundamental ion kinetic scales, producing an unusually steep spectrum just below the break. We present a statistical investigation of the dependence of the break scale on the proton gyroradius $\rho_i$, ion inertial length $d_i$, ion sound radius $\rho_s$, proton-cyclotron resonance scale $\rho_c$ and disruption scale $\lamd$ as a function of $\beta_{\perp i}$. We find that the steepest spectral indices of the dissipation range occur when $\beta_e$ is in the range of 0.1-1 and the break scale is only slightly larger than the ion sound scale (a situation occurring 41\% of the time at 1 AU), in qualitative agreement with the reconnection model. In this range the break scale shows remarkably good correlation with $\lamd$. Our findings suggest that, at least at low $\beta_e$, reconnection may play an important role in the development of the dissipation range turbulent cascade and causes unusually steep (steeper than -3) spectral indices.
\end{abstract}




\section{Introduction}\label{sec:intro2}

Plasma in the solar wind exhibits a turbulent cascade over a very wide range of scales \citep{chen2016}. The turbulent power spectrum consists of several power-law-like ranges, in which different physical mechanisms are involved in the transfer of energy to smaller scales. The scaling behavior of the spectral breaks between these power laws, as well as the spectral indices in the different ranges, provide useful ways to test different turbulence models.

In the ``inertial range", between the outer scale $L_\perp$ and a ``ion kinetic break scale" $\lamb\ll L_\perp$, the turbulence appears to mainly consist of strongly nonlinear, highly anisotropic \citep{horanis}, Alfv\'enically polarized \citep{belcher1971} fluctuations propagating up and down the background magnetic field. The power spectrum of magnetic fluctuations in this range is generally close to $E(k_\perp) \propto k_\perp^{-5/3}$ \citep[e.g.,][]{matthaeus82,chenmallet}.

Below the ion kinetic break scale $\lamb$, in the so-called ``dissipation range", the turbulent spectrum steepens -- generally the spectral index is approximately $-2.8$ or steeper in this range (below $\lamb$ but above a second break or exponential cutoff at electron kinetic scales) \citep{alexandrova2009,sahraoui2010}. This steepening of the spectrum has been explained \citep{schektome2009, howes2011,boldyrevkaw2012} by the fact that below the characteristic ion kinetic scales, the dispersion relation of the characteristic fluctuations of the plasma changes. At moderate-to-high ion plasma beta ($\beta_i=2\mu_0 n_i k_B T_{i}/B_0^2$)\textbf{,} Alfv\'en waves (AW) transition into dispersive kinetic Alfv\'en waves (KAW) when the perpendicular wavenumber becomes comparable to the inverse gyroradius, $k_\perp \rho_i \sim 1$, where $\rho_i=v_{\mathrm{th}\perp i}/\Omega_i$, the ion's perpendicular thermal speed is $v_{\mathrm{th}\perp i}=\sqrt[]{2k_B T_{\perp i}/m_i}$ and the ion gyrofrequency is $\Omega_i =ZeB_0/m_i$. For $\beta_i \ll 1$ (and simultaneously, $\beta_e = 2\mu_0 n_e k_B T_e / B_0^2 \ll 1$), this transition occurs at $k_{\perp} \rho_s \sim 1$, where $\rho_s = \rho_i \sqrt{ZT_e/2T_i} $ is the ion sound radius. Thus, one might expect $\lamb \sim \rho_i$ at $\beta_i \gtrsim 1$ and $\lamb \sim \rho_s$ at $\beta_i \ll 1$. The former scaling appears in measurements of the break scale at $\beta_i \sim 1$ \citep{sahraoui2010,alexandrova2009}. However, \cite{chenbreak2014} studied the behavior of $\lamb$ in two different regimes. For $\beta_i \gg 1$, they found $\lamb \sim \rho_i$ as expected from the KAW dispersion relation. On the other hand, for $\beta_i \ll1$, they found that the break scale was much closer to the ion inertial length, $\lamb \sim d_i =c/\omega_{pi}=\rho_i/\sqrt{\beta_i}$ where $\omega_{pi} = \sqrt[]{n_i Z^2e^2 / \epsilon_0m_{i}}$ is the ion plasma frequency; rather than $\lamb \sim \rho_s$ as would be expected from the KAW dispersion relation. Several studies have suggested that the break frequency in the $\beta_{i} \sim 1$ case can be well approximated with the proton-cyclotron resonance scale defined as $\rho_c = d_i + \sigma_i$ where the pseudo-gyroscale $\sigma_i=v_{th||i}/\Omega_i$ and $v_{th||i}$ is the ion's parallel thermal speed \citep[e.g.][]{bruno2014radial, bruno2015spectral, woodham2018proton}. This method relies on the cyclotron resonance condition for protons, which is satisfied when $k_{||}$ is large enough to allow resonance with the proton population. Since the turbulence is usually highly anisotropic ($k_{\perp} \gg k_{||}$) \citep[e.g.][]{chen2010anisotropy,chen2010interpreting} the measured frequency spectrum generally corresponds to a $k_{\perp}$ wavenumber spectrum and so the proton-cyclotron resonance scale cannot explain the break without also posing an injection of energy into magnetic fluctuations at high $k_{||}$
(e.g. by instabilities, see \cite{klein2015predicted}).

The $\beta_i$-dependent behavior of the ion-scale break is thus somewhat of a mystery. The goal of this Letter is to study the behavior of $\lamb$ across the whole range of $\beta_{\perp i}$ in the solar wind, thus extending the work of \cite{chenbreak2014} on how $\lamb$ behaves at extreme $\beta_{\perp i}$ values.

Recently, \cite{msc_cless} and \cite{loureiroboldyrev_cless} proposed that sheet-like turbulent structures naturally generated by the inertial range turbulence dynamics \citep{boldyrev,Chandran14,mallet3d,howes2016,ms16,verdini20183d} may be disrupted by the onset of reconnection below a characteristic ``disruption scale",
\begin{equation}
\lamd [n=2] = C_D L_\perp^{1/9} (d_e\rho_s)^{4/9},
\label{eqn:ldb}
\end{equation}
where $d_e=c/\omega_{pe}$ is the electron inertial length,  $\omega_{pe}=\sqrt{n_e e^2/\epsilon_0m_e}$ is the electron plasma frequency and $C_D$ is an undetermined dimensionless prefactor of order unity. $\lamd[n=2]$ is the disruption scale of the so-called ``$n=2$" fluctuations, which determine the second-order structure function and power spectrum. Since a detailed study of intermittency (fluctuations with $n\neq2$) is beyond the scope of this paper, we will adopt Equation ~\ref{eqn:ldb} as a single disruption scale, henceforth denoted $\lamd$.

Besides Equation~\ref{eqn:ldb} there is an alternative scaling proposed by \cite{loureiroboldyrev_cless} that relies on a different tearing profile \begin{equation}
\lamd [n=2] \sim L_{\perp} (d_e / L_{\perp})^{8/21} (\rho_s / L_{\perp})^{10/21}.
\label{eqn:LB17}
\end{equation}
The two scalings are observationally indistinguishable from one another in our data set.
Due to this close agreement, we elect to use Equation ~\ref{eqn:ldb} in this Letter.

Equation \ref{eqn:ldb} is only valid when $\lamd$ is larger than the fundamental ion kinetic scale
at which the waves become dispersive (i.e., $\rho_s$), which happens for $\beta_e < \beta_e^{crit}$ given by
\begin{equation}
\beta_{e}^{\rm crit} =C_D^{9/2} \frac{Z m_e}{2 m_i} \left(\frac{L_\perp}{\rho_s}\right)^{1/2},
\label{eqn:critbeta}
\end{equation}
thus, at low $\beta_e$, reconnection may induce a break to a steeper spectrum at a larger scale than one might expect solely on the basis of the KAW dispersion relation. 

This reconnection model relies on the phenomenon of dynamic alignment, which leads to three-dimensionally (3-D) anisotropic eddies and a -3/2 spectral index in the inertial range \citep{boldyrev, Chandran14, ms16}. Although solar wind measurements indicate that the spectral index is closer to -5/3, several observational studies have found clear evidence for 3-D anisotropy of the turbulence \citep{chen3d, vech2016testing,verdini20183d} suggesting that one might expect the structure to be unstable to the onset of reconnection at $\lamd$.

\citet{msc_cless} suggested that the turbulent fluctuations are converted from sheet-like structures above $\lamd$ to flux-rope-like (or vortex-like) structures just below $\lamd$ -- such ``Alfv\'en vortex" structures have been observed in the solar wind just above the ion scales \citep{Alexandrova2008,lion2016,perrone2016,roberts2016observation,perrone2017coherent}, although the exact mechanism generating these structures is a matter of debate.
This significantly accelerates the cascade of the disrupted structures. In order to maintain constant energy flux through scale, the turbulent structures therefore adjust with a sudden drop in amplitude at the disruption scale.
The flux-rope-like structures then cascade as is usual in Alfv\'enic turbulence, becoming progressively more sheet-like, and so on -- until the scale at which the KAW dynamics take over, $\rho_s$.
The relatively shallow spectral index associated with the usual Alfv\'enic dynamics present in this secondary cascade will act to ``smooth out" the rapid drop in amplitude associated with disruption events. This smoothing is increasingly effective as the scale separation between $\lamd$ and $\rho_s$ increases; i.e., as the range of scales over which the usual Alfv\'enic dynamics apply becomes more important relative to the sudden drop in amplitude caused by disruption.
Specifically, \citet{msc_cless} predict that between $\lamd$ and $\rho_s$, the power spectrum would be steeper than $k_\perp^{-3}$, becoming progressively steeper as $\lamd\to\rho_s$ from above (i.e., the spectrum is predicted to be steepest when $\beta_e$ is only moderately low, so that $\lamd$ is only slightly larger than $\rho_s$). 
Therefore, reconnection may fundamentally change the nature of the small scale fluctuations.

In this Letter we use over 13 years of Wind spacecraft data to study the ion spectral break scale $\lamb$  and dissipation-range spectral index, and how these depend on fundamental ion length scales $\rho_i$, $\rho_s$, $d_i$, $\rho_c$, and the disruption scale $\lamd$, as well as on the fundamental parameters $\beta_{\perp i}$ and $\beta_e$. We will show that, at least in terms of scalings, $\lamd$ from the reconnection model \citep{msc_cless} seems to correlate with the measured behavior of the break scale $\lamb$ of the solar wind turbulent power spectrum better than any of the fundamental ion kinetic scales. In addition to this, the steepest spectral indices appear at moderately low $\beta_e$, and when $\lamb$ is only slightly larger than $\rho_s$, as expected qualitatively from the reconnection model. Both of these observations suggest that reconnection may play an important role in the development of the dissipation range turbulent cascade. 

\section{Method} \label{sec:method}
We use a statistical approach based on Wind spacecraft observations to study the variation of the break scale as a function of physical parameters. The investigated period extends from January 2004 to December 2016 during which Wind was in the pristine solar wind. The time series of the magnetic field (11 Hz) \citep{lepping1995wind}, onboard ion moments, ion parameters (92 s cadence both) and electron moments (37 s cadence) \citep{lin1995three, ogilvie1995swe} were split into 10-min intervals ($\sim 5.8 \cdot 10^5$ intervals overall) and the averages of $\beta_{\perp i}$, $\beta_e$, $d_i$, $\rho_i$, $\rho_s$, and $\rho_c$ in each interval were calculated. 

To estimate $\lamd$, we use Equation~\ref{eqn:ldb}, and assume that the break frequency between the energy-containing and inertial ranges is a constant $10^{-4}$ Hz \citep{podesta2007spectral}, calculating $L_{\perp}=V_{sw}/(2\pi 10^{-4})$. The median value of $L_{\perp}$ in our study is $7.4 \cdot 10^5$ km, in good agreement with previous studies,  which suggest that $L_{\perp} \sim 10^6$ km under average solar wind conditions \citep{matthaeus2014nonlinear}. While we do not expect the outer scale to be truly constant over the 13 years of data, in practice this does not introduce a significant source of error in our estimate of $\lamd$, since it appears only as a nearly-constant factor of $L_\perp^{1/9}$ in Equation~\ref{eqn:ldb}. We will determine the dimensionless prefactor $C_D$ in Equation ~\ref{eqn:ldb} from the data.

For each interval, the power spectral density (PSD) of each magnetic field component was computed via Fourier transform and then the components were summed up to obtain the total PSD \citep{koval2013magnetic}. The spectral index and ion-scale break frequency ($\fb$) were identified using the approach of \cite{vech2017nature}. A sequence of 43 logarithmically spaced frequencies was generated from 0.1 to 5.17 Hz and 33 linear fits were made in frequency ranges between the $i$th and $i+10$th elements of this sequence having a ratio of 2.55. From this set of fits, the steepest spectral index was selected and the low frequency end was identified as $\fb$. The overall distribution of $\fb$ shows excellent agreement with the study of \cite{markovskii2008statistical} where $\fb$ was identified manually for 454 solar wind intervals. The mean and standard deviation of the dissipation range spectral index is $-2.99 \pm 0.65$ in excellent agreement with previous studies \citep[e.g.][]{leamon1998observational, smith2006dependence}. We note that we filtered out cases when the dissipation range spectral index and $\fb$ were affected by the noise floor; see \cite{vech2017nature}.

\section{Results} \label{sec:res}
\begin{figure}
    \figurenum{1}
    \centering\includegraphics[width=1\linewidth]{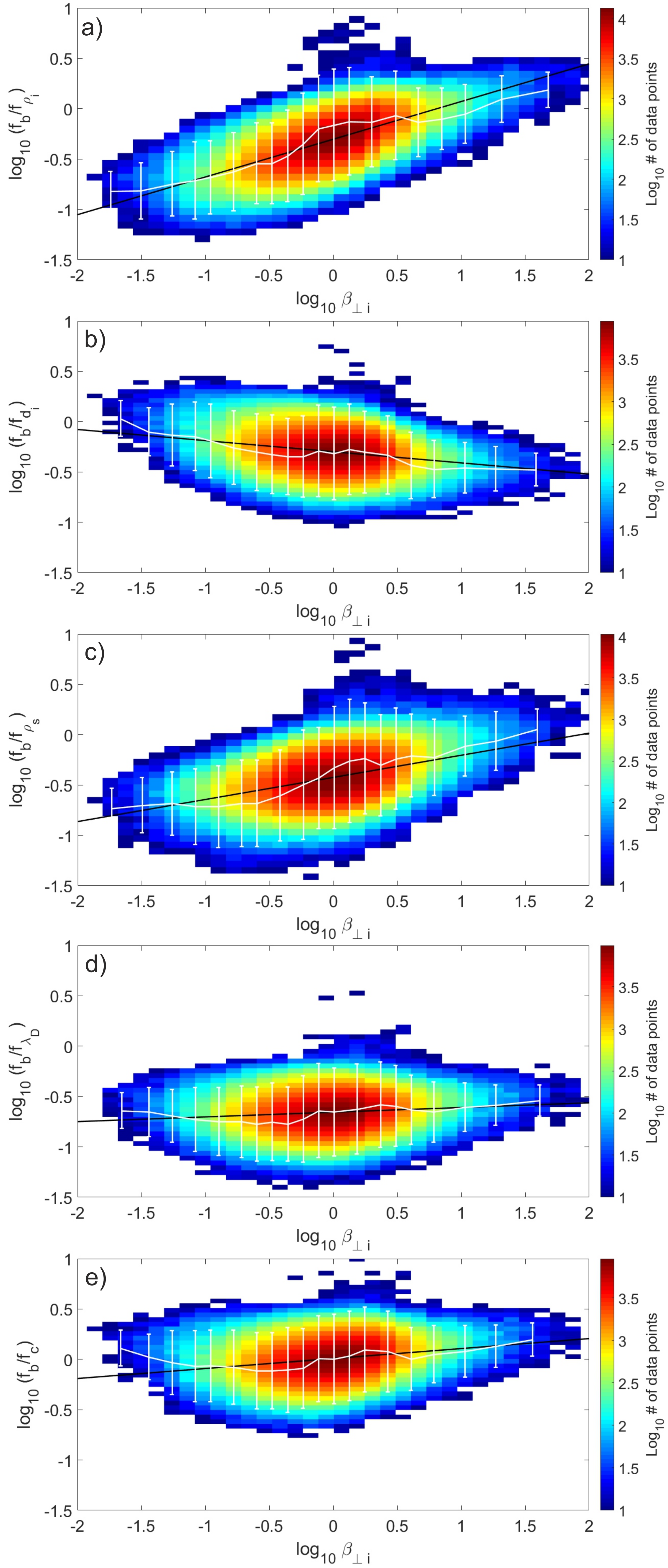}
\caption{The 2-D histograms show the number of data points in each bin in the (a) ($\fb/f_{\rho_i}$, $\beta_{\perp_i}$), (b) ($\fb/f_{d_i}$, $\beta_{\perp_i}$), (c) ($\fb/f_{\rho_s}$, $\beta_{\perp i}$), (d) ($\fb/f_{\lamd}$, $\beta_{\perp i}$) and (e) ($\fb/f_{c}$, $\beta_{\perp i}$) grids, respectively. In each panel, least-square fits are indicated with black lines; their slopes are summarized in Table~\ref{table:tab1}. For each 5\% of the data (as binned by the quantity on the $x$-axis), the averages and standard deviation of the quantity on the $y-$axis are plotted in white.}
  \label{fig:1}
\end{figure}

To study the scaling behavior of $\fb$ as a function of physical parameters in Figure~\ref{fig:1}, we plot 2-D histograms of  $\fb$ normalized to frequencies corresponding to five scales of interest. In all panels, the data is binned in a 50x50 grid, and bins with fewer than 10 samples are discarded. For each 5\% of the data (as binned by the quantity on the $x$-axis), the averages and standard deviation of the quantity on the $y-$axis are plotted, together with the best power law fit to the whole 2-D distribution of the raw data. These power-law exponents and their 95\% confidence intervals from Figure~\ref{fig:1} are summarized in Table~\ref{table:tab1}.

Figures~\ref{fig:1}a and b show the ratio of the break frequency and the frequency corresponding to the convected ion gyroradius ($ f_{\rho_i} =V_{sw} / (2\pi \rho_i$)) and  ion inertial length ($f_{d_i} = V_{sw} / (2\pi d_i$)) as a function of $\beta_{\perp i}$, respectively. Our results agree with \cite{chenbreak2014}: for solar wind intervals with $\beta_{\perp i}  \ll 1$ the break closely aligns with $f_{d_i}$ while in the $\beta_{\perp i}  \gg 1$ case the break is closest to $f_{\rho_i}$: however, this appears to be something of a coincidence -- the white curves and the 2D histograms show no sign of ``flattening" and becoming independent of $\beta_{\perp i}$ at high or low $\beta_{\perp i}$ in Figures~\ref{fig:1}a and b respectively. We therefore, find little evidence that the behavior of the break is truly explained by either $\rho_i$ at $\beta_{\perp i}\gg 1$ or $d_i$ at $\beta_{\perp i}\ll 1$. Indeed, the white lines in Figure~\ref{fig:1} show no significant difference from the overall best fit power laws shown in black for any value of $\beta_{\perp i}$. Overall, the break frequency shows significantly stronger dependence on $f_{\rho_i}$ than $f_{d_i}$: we will discuss one potential reason for the rather shallow dependence of $\fb/f_{d_i}$ at the end of this section.

Figure~\ref{fig:1}c shows the ratio of $\fb/f_{\rho_s}$ ($f_{\rho_s}=V_{sw} / (2\pi \rho_s$)) as a function of $\beta_{\perp i}$, respectively. Similarly to $\fb/f_{\rho_i}$ and $\fb/f_{d_i}$, $\fb/f_{\rho_s}$ has a clear dependence on $\beta_{\perp i}$. We therefore conclude that neither $\rho_i$, $d_i$, nor $\rho_s$ can physically explain the behavior of the ion break scale in the solar wind.

On the other hand, Figures~\ref{fig:1}d shows that $\fb / f_{\lamd}$ ($f_{\lamd}=V_{sw} / (2\pi  \lamd$)) is nearly constant as a function of $\beta_{\perp i}$. This suggests that the ion break scale in the solar wind may be determined by $\lamd$ given by Equation~\ref{eqn:ldb}, as predicted by the reconnection model of \cite{msc_cless}. One obvious caveat to this is that the value of $\fb/f_{\lamd}$ is significantly less than unity, by around half an order of magnitude, across the whole range of $\beta_{\perp i}$. 

In Figure~\ref{fig:1}e, the frequency corresponding to the proton-cyclotron resonance scale ($f_{c}=V_{sw} / (2\pi  \rho_c$)) also shows reasonable agreement with the break frequency across the entire distribution of $\beta_{\perp i}$: its best fit slope parameter is only slightly larger than that of $\fb / f_{\lamd}$ (Table~\ref{table:tab1}). Based on these observations alone, we can not conclusively identify if $\lamd$ or $\rho_c$ controls the break frequency.

\begin{table}
  \caption{Summary of the power-law fits shown in Figure 1,3 and 4} 
  \begin{threeparttable}
    \begin {tabularx}{\linewidth}{c*{3}{>{\hskip0pt}X}}
      \toprule %
      &Parameters & Slope of the power law fit & 95\% confidence interval\\ \toprule %
       & $\fb / f_{\rho_i}$ vs. $\beta_{\perp i}$  & 0.377 & [0.376, 0.379] \\
       & $\fb / f_{d_i}$ vs. $\beta_{\perp i}$  & -0.107 & [-0.109, -0.106] \\
    & $\fb / f_{\rho_s}$ vs. $\beta_{\perp i}$  & 0.219 & [0.2183, 0.221] \\
       & $\fb / f_{\lambda_D}$ vs. $\beta_{\perp i}$  & 0.046 & [0.0454, 0.0479] \\
       & $\fb / f_{c}$ vs. $\beta_{\perp i}$  & 0.099 & [0.0979, 0.1005] \\\midrule
        & $\fb / f_{\lamd}$ vs. $\beta_{\perp i}$  & $-1.9 \times 10^{-4}$ & [-0.0020, 0.0016] \\
        & $\fb / f_{c}$ vs. $\beta_{\perp i}$  & 0.1047 & [0.1028, 0.1066] \\\midrule
       & $\beta_i$ vs. $T_e/T_p$  & -0.317 & [-0.3200, -0.3156] \\
       \midrule
    \end {tabularx}
  \end{threeparttable}
  \label{table:tab1}
\end{table}

\begin{figure*}[ht!]
\figurenum{2} \plotone{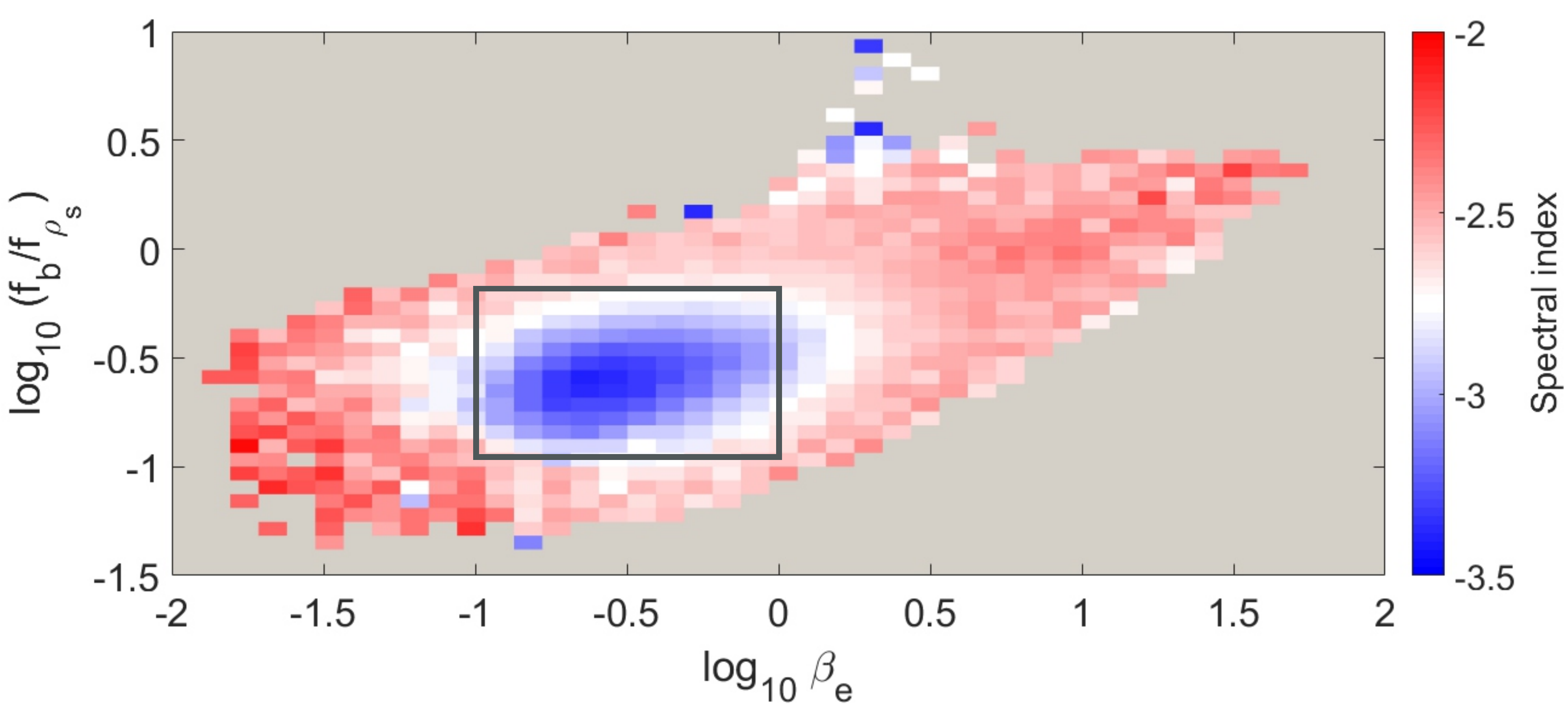}
\caption{Spectral index of the dissipation range binned in the ($\fb/f_{\rho_s}$, $\beta_e$) plane. The black square marks the region with the steepest spectral indices in the range of $0.1\lesssim\beta_e\lesssim 1$ and $0.12 \lesssim \fb/f_{\rho_s} \lesssim 0.63$.
}
  \label{fig:2}
\end{figure*}

\begin{figure}
    \figurenum{3}
    \centering\includegraphics[width=1\linewidth]{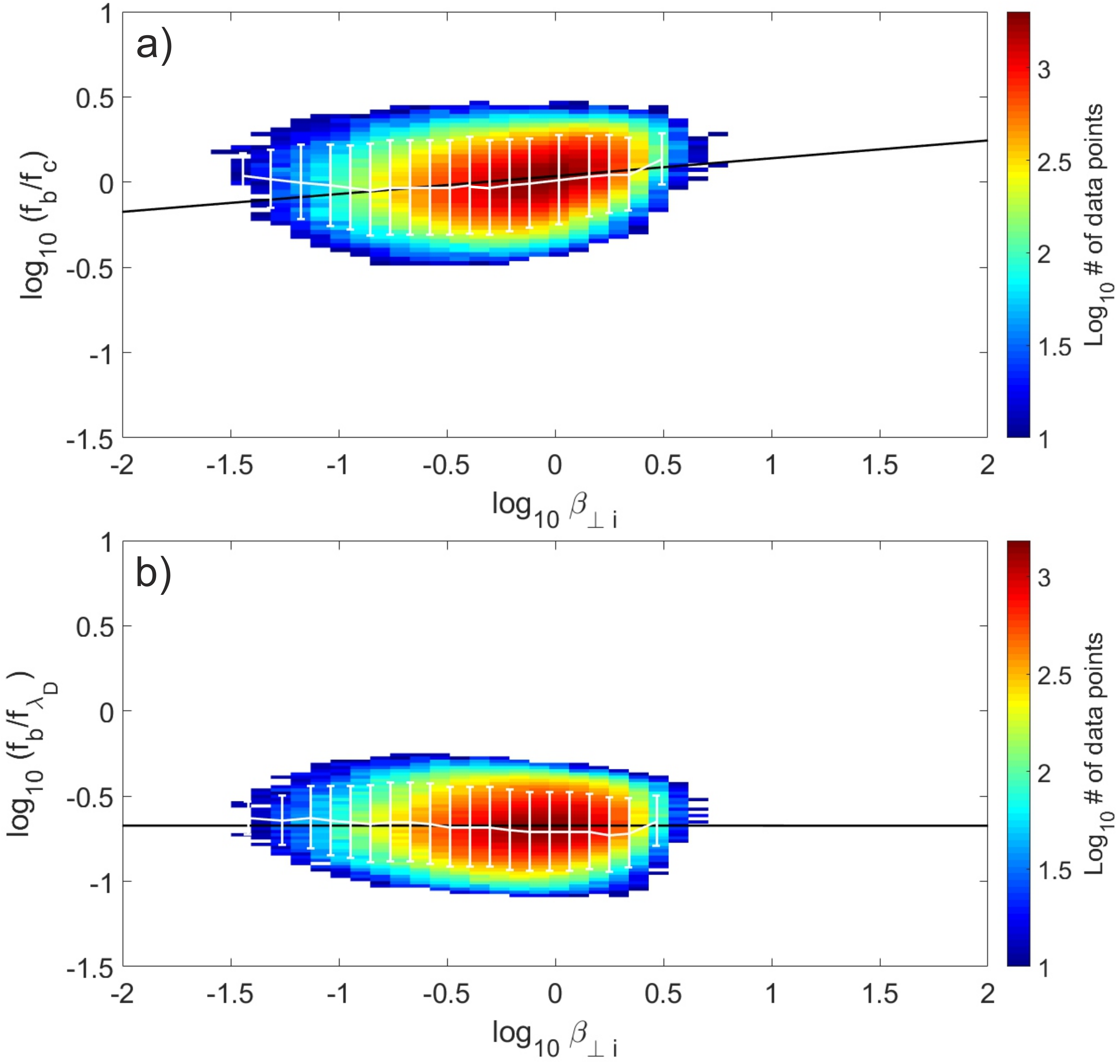}
\caption{a) and b) are identical to Figure~\ref{fig:1}e and d, however they present the subset of measurements, which are within the black square in Figure~\ref{fig:2} corresponding to 41\% of the overall data points.}
  \label{fig:3}
\end{figure}

\begin{figure}
    \figurenum{4}
    \centering\includegraphics[width=1\linewidth]{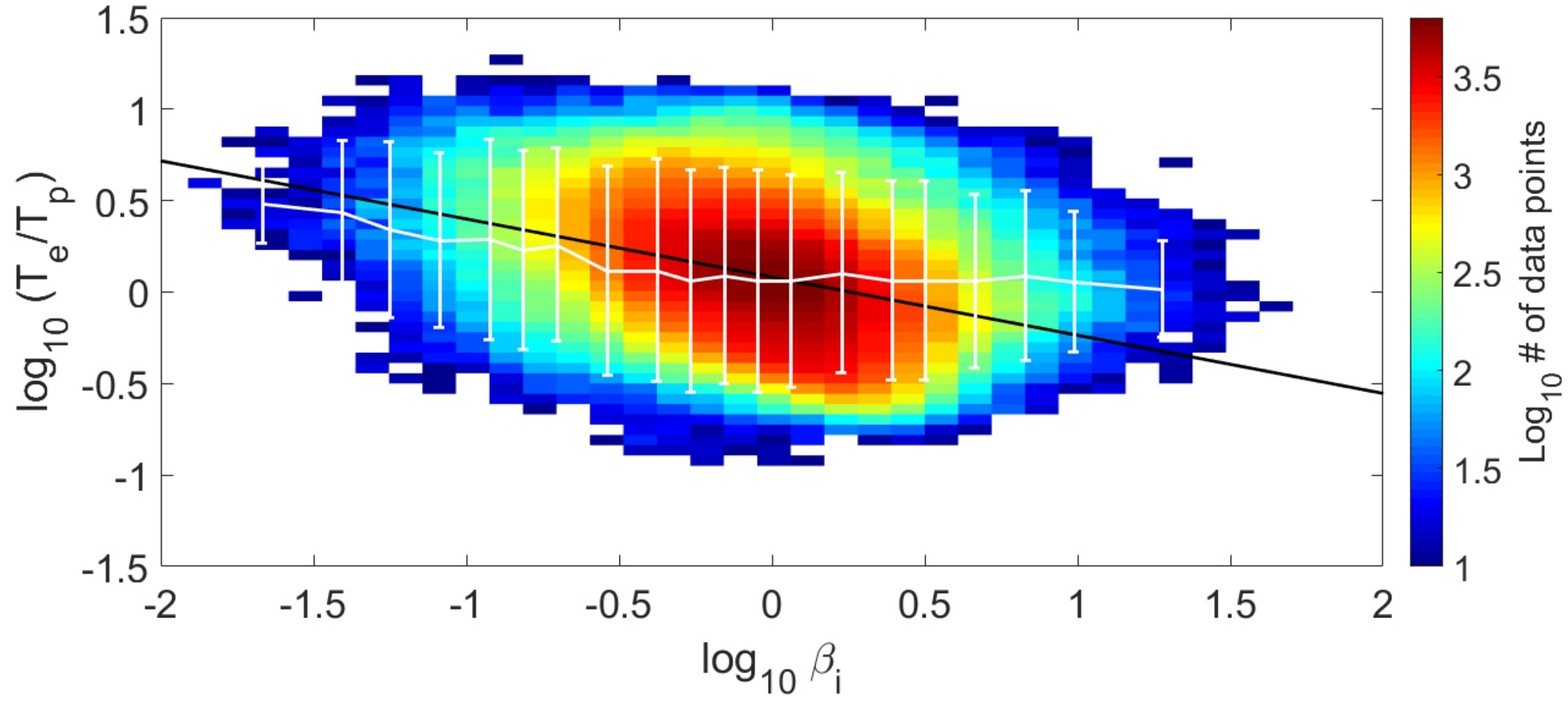}
\caption{2-D histogram showing the distribution of the measurements in the ($\beta_i,T_e/T_p$) plane. The least-square fit is indicated with a black line and the average and standard deviation of each 5\% of the data are marked in white.}
  \label{fig:4}
\end{figure}

In Figure~\ref{fig:2} the ($\fb/f_{\rho_s}, \beta_e$) plane is shown and the color represents the median dissipation range spectral index in each bin. The distribution indicates a significant steepening of the spectral index at ``moderately" small $\beta_e$ values between approximately 0.1 - 1, where $\fb$ is slightly smaller than $f_{\rho_s}$. In this region the spectral indices are typically steeper than -3 in a narrow range just above the break \citep[cf.][]{sahraoui2010}. \citet{msc_cless} predict that the steepest indices should be attained for $\beta_e$ values just low enough that the reconnection-induced break occurs only just before the transition from AW to KAW (at the ion scale). Our technique (see Section 2) measures the spectral index over a fixed range of $[f_b, 2.55\cdot f_b]$. If there is a steep subrange narrower than this just above the break, this approach cannot capture its true steepness. Due to this limitation in our fitting technique as well as the narrow range of scales involved, we are only able to claim qualitative similarity with the \citet{msc_cless} prediction.

To resolve the ambiguous results obtained with Figure~\ref{fig:1}d and e, we repeat our analysis with the subset of the data within the black square in Figure~\ref{fig:2}, which encloses 41 \% of the total data and is bounded by $0.1\lesssim\beta_e\lesssim 1$ and $0.12 \lesssim \fb/f_{\rho_s} \lesssim 0.63$. The spectral indices are significantly steeper in this region, and if this is caused by reconnection then $\lamb$ may have significantly better scaling with $\lamd$ in the marked region than with $\rho_c$. In Figure~\ref{fig:3}a the slope of the power-law fit for $\fb/f_{c}$ as a function of $\beta_{\perp i}$ closely agrees with the one based on the full distribution. In contrast, in Figure~\ref{fig:3}b the slope of the power-law fit for $\fb/f_{\lamd}$ as a function of $\beta_{\perp i}$ is remarkably close to 0. Based on this, we suggest that at least at low $\beta_e$ magnetic reconnection may control the ion-scale break of the solar wind turbulence. The intercept of the power-law fit is $10^{-0.6428}$, implying $C_D = 4.7$ (see Equation ~\ref{eqn:ldb}). 

Both $\fb/f_{\lamd}$ and $\fb/f_{c}$ were close to normally distributed in our data thus we use an F-test to investigate whether they have equal variance. The ratio of the sample variances is $F = \sigma_{(\fb/f_{\lamd})}^2 / \sigma_{(\fb/f_c)}^2 =0.8398$ with 95 \% confidence intervals of [0.8328; 0.8468]. Thus we reject the null hypothesis and conclude that $\lamd$ predicts the break with smaller spread than $\rho_c$.

Finally, the scaling of $\fb/f_{d_i}$ as a function of $\beta_i$, while significant over the whole range of $\beta_i$ present in the data, is not particularly strong. With reference to the results of \citet{chenbreak2014}, \citet{msc_cless} pointed out that correlations between $T_e/T_i$ and $\beta_i$ could cause the scaling behavior of $d_i$ to mimic that of $\lamd$. Neglecting a nearly constant factor $(L_\perp/\rho_s)^{1/9}$,
\begin{equation}
\frac{\lamd}{\rho_s}\propto \beta_e^{-2/9} \Rightarrow \frac{\lamd}{d_i} \propto \left(\beta_i \frac{T_e}{T_i}\right)^{5/18},
\end{equation}
and so if $T_e/T_i$ were anticorrelated with $\beta_i$, the ratio $\lamd/d_i$ would scale less strongly with $\beta_i$ than otherwise expected. Figure~\ref{fig:4} shows that empirically, such an anticorrelation does in fact exist in the solar wind (at least at low $\beta_i$). Assuming that the break scale $\lamb \propto \lamd$ (as does appear to be the case: see Figures~\ref{fig:1}d and ~\ref{fig:3}b), this contributes to the rather shallow scaling of $\fb/f_{d_i}$ in Figure~\ref{fig:1}b. Similar consideration could contribute to the shallow scaling of $\fb/f_{c}$ with $\beta_{\perp i}$.

\section{Conclusion} \label{sec:conclusion}
In this Letter we have presented a statistical study of the break scale $\lamb$ between the inertial and dissipation ranges of the solar wind turbulence spectrum, and to what extent $\lamb$ agrees with the fundamental ion length scales $\rho_i$, $d_i$, $\rho_s$, $\rho_c$ and the disruption scale $\lamd$ (Equation ~\ref{eqn:ldb}), as a function of $\beta_{\perp i}$. Our results suggest that the ion-scale break of the solar wind turbulence may be controlled by magnetic reconnection in the low $\beta_e$ (0.1-1) case, which is the main result of this Letter.

The observed behavior of $\fb/f_{\rho_i}$ and $\fb/f_{d_i}$ as a function of $\beta_{\perp i}$ are consistent with previous studies based on more limited data sets \citep{chenbreak2014,wangion2018}: for $\beta_{\perp i } \ll 1$ the break occurs at the frequency of $f_{d_i}$ and for $\beta_{\perp i} \gg 1$ the break is closest to $f_{\rho_i}$. 
However, both $\fb/f_{\rho_i}$ and $\fb/f_{d_i}$ showed significant dependence on $\beta_{\perp i}$ across the whole range of $\beta_{\perp i}$ present in the data, suggesting that the agreement with the break frequency in narrow ranges at the extremes of $\beta_{\perp i}$ is somewhat coincidental. Thus, we find little evidence that either $\rho_i$ or $d_i$ determine the break in the power spectrum. Similarly, $\fb/f_{\rho_s}$ has a clear scaling with $\beta_{\perp i}$ and thus cannot explain the position of the break. 

This contrasts with recent hybrid simulations by \citet{franci2016plasma} which found that for $\beta_{\perp i}\ll 1$, $\lamb \sim d_i$ independently of $\beta_i$, and for $\beta_{\perp i}\gg 1$, $\lamb \sim \rho_i$ independently of $\beta_{\perp i}$. We note, however, that their simulations may contain different physics than the true solar wind turbulence; they are two-dimensional, and also do not contain the electron inertial scale, which allows the reconnection to occur in the model of \cite{msc_cless}.

Comparing the break scale to the scale predicted by the reconnection model \citep{msc_cless,loureiroboldyrev_cless}, we found that $\fb/f_{\lamd}$ was nearly independent of $\beta_{\perp i}$. To obtain agreement in the magnitudes of $\lamb$ and $\lamd$ a dimensionless prefactor $C_D = 4.7$ must be inserted into Equation~\ref{eqn:ldb} since this cannot be predicted from the simplified model in \cite{msc_cless}, which only predicts scalings. The ratio of the proton-cyclotron resonance scale $\rho_c$ to the break scale showed a similarly shallow scaling with $\beta_{\perp i}$ and the best fit-slope was only a factor of 2.15 steeper than that of $\fb/f_{\lamd}$. Thus based on the whole distribution of the data we can not conclusively identify if $\lamd$ or $\rho_c$ controls the break frequency. 

At high $\beta_e$, reconnection is not expected to cause a break (see Equation ~\ref{eqn:critbeta}); this could be why the agreement between $\lamb$ and $\lamd$ has a slight dependence on $\beta_{\perp i}$ using the whole dataset.
We therefore repeated our analysis for a significant subset of the data (41 \% overall) bounded by $0.1\lesssim\beta_e\lesssim 1$ and $0.12 \lesssim \fb/f_{\rho_s} \lesssim 0.63$ displaying unusually steep spectral indices (steeper than -3), in qualitative agreement with the reconnection model of \cite{msc_cless}. For this subset of the data we found that $\lamb$ scales with $\lamd$ remarkably well while $\beta_{\perp i}$ changes two orders of magnitude. In contrast, $\rho_c$ showed a clear correlation with $\beta_{\perp i}$ suggesting that at least at the low $\beta_e$ case the break between the inertial and dissipation-range scales may be controlled by the onset of magnetic reconnection.

\acknowledgments
A. Mallet was supported by by NSF grant AGS-1624501. K.G. Klein was supported by NASA grant NNX16AM23G. J.C. Kasper was supported by NASA grant NNX14AR78G. Data were sourced from CDAWeb (http://cdaweb.gsfc.nasa.gov/).

\end{document}